# Thermal Expansion and Magnetostriction of the Ising Antiferromagnet TbNi$_2$Ge$_2$

G.M. Schmiedeshoff*, S.M. Hollen*, S.L. Bud'ko¶, and P.C. Canfield¶

*Department of Physics, Occidental College, Los Angeles, CA 90041, USA
¶Ames Laboratory and Department of Physics, Iowa State University, Ames, IA 50011, USA

**Abstract.** We have measured the linear thermal expansion and magnetostriction of the Ising antiferromagnet TbNi$_2$Ge$_2$ along its c-axis from room temperature to 2 K and in magnetic fields to 14 T. We find a magnetic phase diagram that agrees with earlier work and estimate aspects of its uniaxial pressure dependence. We also find a new high field feature near 10 T which may signal the onset of an additional field-induced phase.

**Keywords:** Thermal Expansion, Magnetostriction, Antiferromagnetism.

## INTRODUCTION

The ternary rare-earth intermetallic compound TbNi$_2$Ge$_2$ crystallizes in a body-centered tetragonal ThCr$_2$Si$_2$ structure and exhibits Ising-like antiferromagnetism with phase transitions into incommensurate and commensurate states at $T_N = 16.7$ K and $T_t = 9.6$ K respectively; six additional metamagnetic phases have been observed in applied fields at 2 K. The magnetic properties of TbNi$_2$Ge$_2$ are driven by indirect exchange interactions (presumably of the RKKY variety) between the magnetic rare-earth ions via the conductions electrons. TbNi$_2$Ge$_2$ has been studied by Bud'ko *et al.* as part of a more general inquiry into the magnetic behavior of rare-earth-Ni$_2$Ge$_2$ compounds [1]; references to earlier work may also be found in Ref. [1].

We have measured the linear thermal expansion and magnetostriction along the *c*-axis of TbNi$_2$Ge$_2$ (the axis along which the spins align) using a capacitive dilatometer constructed of OFHC copper and designed to operate in a Physical Property Measurement System available from Quantum Design Inc. A detailed description of the dilatometer will appear elsewhere [2].

## RESULTS AND DISCUSSION

The linear thermal expansion of TbNi$_2$Ge$_2$ along its *c*-axis ($\alpha = d\ln(L)/dT$ where L is the length of the sample, 0.85 mm here) is shown in Fig. 1. Features associated with the phase transitions at $T_N$ and $T_t$ are clearly visible. The broad maximum near 30 K is due to crystal field effects.

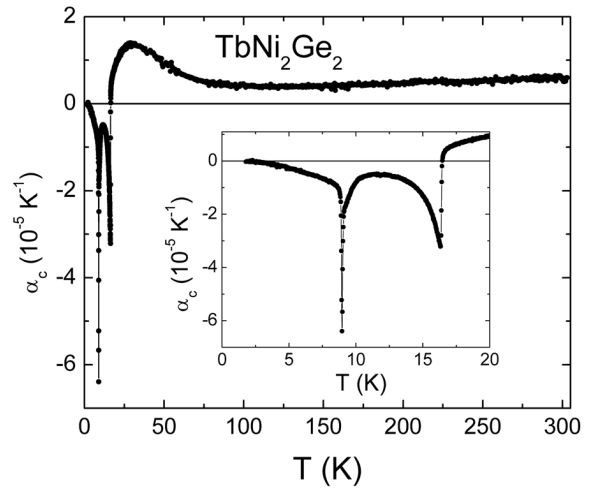

**FIGURE 1.** Linear thermal expansion of TbNi$_2$Ge$_2$ along its c-axis. The inset shows an expanded view of the low temperature phase transitions.

We first focus on the feature at $T_N$, the shape of which suggests that it is due to a 2$^{nd}$ order phase transition. Combining the jump in thermal expansion ($\Delta\alpha = 2.9 \times 10^{-5}$ K$^{-1}$) with the jump in molar specific heat from Ref. [1] ($\Delta C_p = -15$ J/mol K), we can use the Ehrenfest relation

$$\frac{dT_N}{dP_c} = V_M T_N \frac{\Delta\alpha_c}{\Delta C_p}, \qquad (1)$$

where $P_c$ is the uniaxial pressure along the c-axis and $V_M$ is the molar volume, to estimate the uniaxial pressure dependence of $T_N$ to be –0.16 K/kbar.

The sharp feature at $T_t$ suggests that this phase transition is 1st order, a suggestion that is supported by the presence of magnetic hysteresis across this phase boundary in higher magnetic fields [1]. However, the feature observed in $C_p$ at this temperature is relatively broad and therefore at odds with this suggestion. Should this be a 1st order transition, a Clausius-Clapyeron equation would apply:

$$\frac{dT_t}{dP_c} \approx V_M \frac{\Delta(\Delta L/L)}{\Delta S}, \qquad (2)$$

where $\Delta(\Delta L/L)$ is the jump in the expansivity $\Delta L/L = L(T)-L(0)/L(0)$ at the phase transition ($\Delta L/L$ is determined by integrating $\alpha(T)/T$ across the transition), and $\Delta S$ is the jump in molar entropy at the phase transition (determined by integrating $C_p(T)/T$ across the transition). Integrating our data shows that $\Delta(\Delta L/L)$ is clearly finite at $T_t$, but integrating the specific heat data suggests that $\Delta S$ is very close to zero. If this transition is 1st order, Eqn. (1) suggests that the uniaxial pressure dependence of $T_t$ might be very large.

The magnetostriction of $TbNi_2Ge_2$ at 2 K, measured along its c-axis with magnetic fields applied parallel to the c-axis, is shown in Fig. 2 where $\Delta L/L = L(H)-L(0)/L(0)$. Magnetic hysteresis is observed in low fields along with a series of metamagnetic phase transitions consistent with the measurements of Bud'ko et al. [1] up to about 6 T.

We observe an abrupt change in the slope of $\Delta L/L$ near 10 T at 2 K (marked by an arrow in Fig. 2, the derivative of the data with respect to magnetic field in the vicinity of 10 T is shown in the inset of Fig. 2). We suspect that this feature may signal a new field-induced phase transition although there is no corresponding signal of such a transition in high field magnetization measurements [1]. The feature was quite reproducible in our measurements and increased in field as the temperature increased (reaching 12.8 T at 30 K; the feature was not observed at 50 K up to our maximum field of 14 T). This temperature dependence suggests that the feature may be due to a spin-flop-like transition, though one would expect to see an associated metamagnetic-like feature in the field dependent magnetization, a feature that is not observed [1]. Understanding the origin and nature of this feature will require further study.

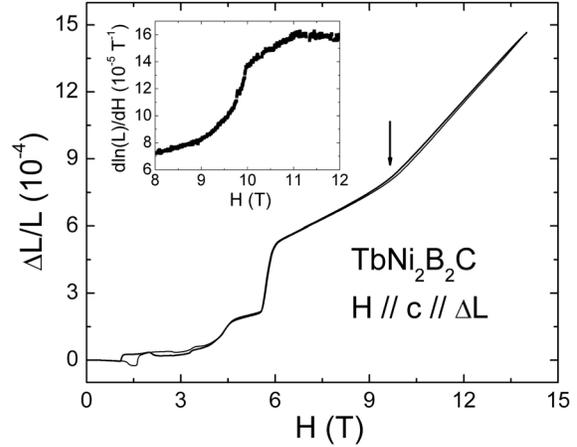

**FIGURE 2.** Magnetostriction of $TbNi_2Ge_2$ at 2 K, measured along its c-axis in magnetic fields applied parallel to the c-axis. The arrow denotes the position of a new feature. The inset shows a derivative of the data with respect to magnetic field in the vicinity of the feature.

## CONCLUSIONS

We have measured the thermal expansion and magnetostriction of $TbNi_2Ge_2$ along its c-axis from room temperature to 2 K and in magnetic fields to 14 T. Our results on the magnetic phase diagram agree with earlier results except for the observation of what may be a new field-induced phase above 10 T at low temperature. Application of thermodynamic relations to our results allowed us to estimate the affect of uniaxial pressure on the magnetic phase transitions in zero field.

## ACKNOWLEDGMENTS

We are grateful to A.H. Lacerda and J.C. Cooley for several helpful discussions. Ames Laboratory is operated for the U.S. Department of Energy by Iowa State University under Contract No. W-7405-Eng-82. This work was supported by the Director of Energy Research, Office of Basic Energy Sciences and by the National Science Foundation under DMR-0305397.